\documentclass[letterpaper]{article} % DO NOT CHANGE THIS
\usepackage{aaai2026} 
\usepackage{times}  % DO NOT CHANGE THIS
\usepackage{helvet}  % DO NOT CHANGE THIS
\usepackage{courier}  % DO NOT CHANGE THIS
\usepackage[hyphens]{url}  % DO NOT CHANGE THIS
\usepackage{graphicx} % DO NOT CHANGE THIS
\usepackage{natbib}  % DO NOT CHANGE THIS AND DO NOT ADD ANY OPTIONS TO IT
\usepackage{caption} % DO NOT CHANGE THIS AND DO NOT ADD ANY OPTIONS TO IT
\usepackage{algorithm}
\usepackage{algorithmic}

\usepackage{newfloat}
\usepackage{listings}
\DeclareCaptionStyle{ruled}{labelfont=normalfont,labelsep=colon,strut=off} % DO NOT CHANGE THIS
\floatstyle{ruled}
\newfloat{listing}{tb}{lst}{}
\floatname{listing}{Listing}

\usepackage{booktabs}
\usepackage{xcolor}

\usepackage{enumitem}

\newcommand{\new}[1]{{#1}}

\usepackage{multirow}

\newcommand{\answerYes}[1]{\textcolor{blue}{#1}} 
\newcommand{\answerNo}[1]{\textcolor{teal}{#1}} 
\newcommand{\answerNA}[1]{\textcolor{gray}{#1}}

\usepackage{xspace}
\newcommand{\hide}[1]{}
\newcommand{\eg}{\textit{e.g.,}\xspace}

\newcommand{\vs}{\textit{vs.}\xspace}
\newcommand{\fig}{Figure\xspace}
\newcommand{\sect}{\S}

\title{How Conversational Structure and Style Shape Online Community Experiences}

\author {
    Galen Weld\textsuperscript{\rm 1,2},
    Carl Pearson\textsuperscript{\rm 1},
    Bradley Spahn\textsuperscript{\rm 1}\footnote{Work completed while at Reddit. This author has since moved to OpenAI.},
    Tim Althoff\textsuperscript{\rm 2},
    Amy X. Zhang\textsuperscript{\rm 2},
    Sanjay Kairam\textsuperscript{\rm 1}\footnotemark[1]
}
\affiliations {
    \textsuperscript{\rm 1}Reddit, Inc.\\
    \textsuperscript{\rm 2}University of Washington \\
    {gweld@cs.washington.edu, carl.pearson@reddit.com, bradleyspahn@gmail.com, \\ althoff@cs.washington.edu, axz@cs.washington.edu, sanjay.kairam@gmail.com}
}

\begin{document}

\maketitle

\begin{abstract}
Sense of Community (SOC) is vital to individual and collective well-being. Although social interactions have moved increasingly online, still little is known about the specific relationships between the nature of these interactions and Sense of Virtual Community (SOVC). This study addresses this gap by exploring how conversational structure and linguistic style predict SOVC in online communities, using a large-scale survey of 2,826 Reddit users across 281 varied subreddits. We develop a hierarchical model to predict self-reported SOVC based on automatically quantifiable and highly generalizable features that are agnostic to community topic and that describe both individual users and entire communities.
We identify specific interaction patterns (\eg reciprocal reply chains, use of prosocial language) associated with stronger communities and identify three primary dimensions of SOVC within Reddit -- Membership \& Belonging, Cooperation \& Shared Values, and Connection \& Influence.
% Our hierarchical model explains 33.1\% of the variance in self-reported SOVC using topic-agnostic features that capture the structure and style of social interactions.
This study provides the first quantitative evidence linking patterns of social interaction to SOVC and highlights actionable strategies for fostering stronger community attachment, using an approach that can generalize readily across community topics, languages, and platforms. These insights offer theoretical implications for the study of online communities and practical suggestions for the design of features to help more individuals experience the positive benefits of online community participation.


\end{abstract}
% for author notes
% \footnotetext[]{Work completed while at Reddit. This author has since moved to OpenAI.}

\section{Introduction}\label{sec:intro}
Sense of Community, the bond between individuals and their social groups, plays a critical role in individual and collective well-being \cite{mcmillanchavis1986}. Stronger community attachment, both offline and online, is predictive of increased individual well-being~\cite{farrell2004neighborhoods, kennett2005understanding, sum2009internet}, resilience against stress and high-risk behaviors~\cite{cowman2004mediating, ferrari2007eldercare, welbourne2009supportive, david2016comprehensive, dubois2011mentoring, resnick2004youth, farrell2004neighborhoods, hurd2009negative}, and community resilience and problem-solving~\cite{ahlbrandt1979new, burroughs1998psychological, bachrach1985coping}. 

As billions of people have moved their social interactions into online spaces, understanding the dynamics of Sense of Virtual Community (SOVC) is essential to fostering these positive effects within online environments \cite{koh2003sense}. Prior work has relied primarily on surveys to understand SOVC across listservs and newsgroups~\cite{blanchard2004experienced, blanchard2008testing}, blogs~\cite{blanchard2004blogs, flensted2011exploring, hopkins2010psychological}, discussion forums~\cite{wasko2000participate, welbourne2009supportive, tonteri2011antecedents, abfalter2012sovc, gibbs2019investigating, smith2022governance}, % removed koh2003sense from this to save on space
and livestreaming communities~\cite{kairam2022social}. These studies offer valuable insight into how SOVC manifests but have largely overlooked how specific aspects of community behavior shape these perceptions. This gap limits our understanding of how community behaviors translate into the feelings of belonging, cooperation, and connection that characterize SOVC.

Our study bridges this gap by examining how SOVC manifests across diverse communities on Reddit, as well as how conversational structure and linguistic style predict SOVC. Conversational structure captures the patterns and dynamics of user interactions, such as thread depth, reciprocal exchanges, and voting behaviors~\cite{aragon2017conversations, backstrom2013characterizing, choi2015characterizing, kumar2010dynamics}. These structural elements can predict conversation- or thread-level outcomes, such as engagement and virality~\cite{buntain2014identifying, saveski2021structure, coletto2017automatic}. % removed medvedev2019modelling to save space

In contrast, linguistic style captures the expressive and emotional characteristics of user communication, such as who are the chosen targets of speech acts, whether speech is past- or future-oriented, or whether speakers use emotional or socially-oriented language. Prior work using LIWC (Linguistic Inquiry \& Word Count), a well-validated dictionary-based tool for grouping words into `psychologically meaningful categories'~\cite{tausczik2010psychological}, has shown how linguistic, content-agnostic signals like affect or cognitive processes can predict various outcomes in online communities~\cite{althoff2014ask, dong2023we, matthews2015they, mcewan2016communication, ashokkumar2022tracking, cassell2005language}. Providing insight into the tone and intent of user interactions, these stylistic markers offer a complementary lens to the structural perspective. No prior work has sought to understand how conversational style and structure relate to the broader community-level construct of SOVC.

In this study, we evaluate how community behavior, as measured by both conversational structure and linguistic style categories of features, can predict differences in the SOVC experienced by members of Reddit communities. Reddit's diverse communities, which span a broad range of topics, provide an ideal context for exploring these relationships. Specifically, we address two key research questions:

\begin{itemize}[label={}, leftmargin=*, align=left]
    \item[\textbf{RQ1}] What are the dimensions along which SOVC varies across communities on Reddit (\sect\ref{sec:results_sovc_factors})?
    \item[\textbf{RQ2}] How do patterns of conversational structure and linguistic style explain differences in SOVC between and within communities (\sect\ref{sec:results_modeling})?
\end{itemize}

To answer these questions, we combined survey responses from 2,826 Reddit users with detailed behavioral data capturing interactions within 281 communities.  We answer the first research question using exploratory factor analysis to identify meaningful dimensions which describe how the community experience varies across communities (\sect\ref{sec:methods_features}). We answer the second question by constructing a hierarchical model to predict individual-level and community-level scores for each of these dimensions, based on behavioral trace data (\sect\ref{sec:methods_modeling}) which characterize these communities in terms of their conversational structure and linguistic style.
\new{We intentionally choose automatically quantifiable features which are minimally related to communities' topic and content in order to maximize the degree to which our findings generalize across communities of different topics, languages, and potentially different platforms.}

\subsubsection{Contributions} 
This paper provides the following contributions to the area of computer-mediated communication:
\begin{itemize}
    \item We provide the first large-scale study of SOVC in Reddit, finding that SOVC varies along three dimensions (\textit{Membership/Belonging}, \textit{Cooperation/Shared Values}, and \textit{Connection/Influence}) \new{which we relate to existing research} (\sect\ref{sec:results_sovc_factors}).
    \item We provide the first quantitative evidence linking both conversational structure and linguistic style to differences in SOVC (\sect\ref{sec:results_modeling}). These \new{generalizable} features capture a substantial fraction of the variance of SOVC in our hierarchical model, highlighting the extent to which these patterns of interaction shape the subjective experience of online communities.
    \item We identify actionable features, such as deeper reply chains and prosocial language, that platforms may be able to leverage to foster engagement, improve moderation, and strengthen community bonds (\sect\ref{sec:discussion}).    
\end{itemize}

\new{By modeling the relationship between automatically measurable conversational features and SOVC}, this work lays the foundation for future studies aimed at enhancing online community health and resilience. \new{Our approach enables the study of SOVC across diverse communities and platforms by deliberately excluding features that capture the topic of discussions}, and our methods could extend to non-English and smaller online communities. Our findings open new avenues for research into how behavior patterns shape -- and are shaped by -- social dynamics and offer practical opportunities for improving community experiences, and their associated benefits, at scale.

\section{Related Work}\label{sec:related}

\subsubsection{Predicting SOVC in Online Communities.}
Extensive prior research on Sense of Virtual Community (SOVC) has identified a set of core dimensions that overlap with and extend those identified in studies of offline communities. In study of offline communities, \citet{mcmillanchavis1986} initially proposed a definition of `Sense of Community' (SOC) encompassing four dimensions: \textit{membership} (feeling of belonging or personal relatedness), \textit{influence} (a sense of mattering to the group or other members), \textit{integration and fulfillment of needs} (a feeling that members will provide for each other), and \textit{shared emotional connection} (a sense of shared history and experiences). 

Online communities afford mechanisms for interaction that can produce different styles of community attachment. Prior studies of SOVC have found that, while the concept of membership translates well to online communities~\cite{abfalter2012sovc, blanchard2007developing, blanchard2008testing, blanchard2004blogs, koh2003sense, tonteri2011antecedents, kairam2022social}, SOVC can also manifest along different dimensions, such as immersion~\cite{koh2003sense}, recognition~\cite{blanchard2004blogs}, identification~\cite{blanchard2007developing, blanchard2004blogs, tonteri2011antecedents}, emotional feelings~\cite{blanchard2007developing, tonteri2011antecedents}, or cohesion~\cite{kairam2022social}.

\new{Previous research has examined how activity patterns in online communities relate to Sense of Virtual Community (SOVC).}
\new{
Studies across a broad range of contexts, such as online health forums, online games, and livestreaming communities, have demonstrated strong links between the volume and types of participation and the feelings of SOVC that members experience~\cite{cummings2002beyond, wu2017inferring, kairam2022social, turkay2019friending, Israeli2023WithFC}.}
\new{
In the specific context of Reddit, survey and interview studies have identified factors like interaction volume and quality, member tenure, norm enforcement, and bot usage as factors that influence members' experiences \cite{smith2022governance, prinster2024community, weld2024taxonomy} and examined how these factors vary between different communities \cite{weld2021better}.} \new{However, these studies have not quantified SOVC nor directly connected such factors to measured SOVC,} which is the primary contribution of this work.

\subsubsection{Markers of Linguistic Style.}
The conversations which make up online communities exhibit distinctive stylistic features that prior work has linked to differences in individual and collective outcomes. Many studies have relied on LIWC (Linguistic Inquiry and Word Count), a dictionary-based method for grouping words into `psychologically meaningful' categories \cite{tausczik2010psychological}. \new{LIWC can be used to identify patterns of linguistic style that can be used to reliably differentiate content from different communities, even if they share topics \cite{khalid2020style}, and to profile the distribution of individual personalities or values across a community \cite{kumar2018inducing, Srivastava_2015_Enculturation}. }
\new{While other methods have been used to measure how relative differences in language usage between individuals and community pairs \cite{danescu2013country, israeli_2022_unsupervised_discovery, Waller2020QuantifyingSO, Weld2024PerceptionsOM}, we focus on absolute measurements of language, and we use LIWC for its robustness and generalizability.}

Variation in specific categories within LIWC, such as words expressing group affiliation and those expressing positive/negative emotion, correlates with differences in individual and collective outcomes, including individual retention \cite{ma2017write}, subjective well-being~\cite{dong2023we}, community satisfaction~\cite{matthews2015they}, and community stability~\cite{mcewan2016communication}. A small number of studies have applied LIWC specifically to understanding group identity and cohesion \cite{ashokkumar2022tracking, cassell2005language}. 
In this work, we directly use LIWC features to model SOVC across a diverse set of communities and members.

\subsubsection{Conversational Structure.}
Modeling the structure of online conversations -- as threads, trees, or interaction networks --  reveals important patterns of user engagement, influence and interaction. Key metrics include conversation volume (e.g. total comments)~\cite{choi2015characterizing, kumar2010dynamics}, depth (reply chain length)~\cite{choi2015characterizing, kumar2010dynamics, yu2024deconstructing}, breadth (reply chain branching)~\cite{yu2024deconstructing}, and degree distribution~\cite{choi2015characterizing, kumar2010dynamics, yu2024deconstructing}. Network-based approaches have also modeled conversational dynamics to understand influence and community formation~\cite{buntain2014identifying, kou2018understanding, saveski2021structure}.

A particularly useful approach involves analyzing local interaction patterns, such as recurring motifs within conversation graphs. \new{\citet{coletto2017automatic} used network motifs -- small, recurring subgraphs of user interactions -- to detect controversial discussions, finding that specific dyadic and triadic patterns reliably identify divisive exchanges. \citet{paranjape2017motifs} introduced temporal motifs to capture dynamic elements, such as response delays, that static models often miss.} \new{\citet{kuvsen2021building} used temporal motifs to differentiate controversial and non-controversial exchanges. \citet{zhang2018characterizing} expanded on structural motifs to include engagement signals (\eg likes), an approach that we leverage in the present study.}

These structural features have been successfully used across platforms to predict various conversational outcomes without content signals. Examples include forecasting discussion thread lengths on Facebook and Wikipedia~\cite{backstrom2013characterizing} and on Reddit~\cite{horawalavithana2022online}, and modeling the controversiality~\cite{coletto2017automatic}, toxicity~\cite{saveski2021structure}, and emotional tenor~\cite{kuvsen2021building} of Twitter conversations. \new{\citet{zhang2018characterizing} present the prior work most similar to our own, using conversational motifs to characterize communities along various dimensions (e.g. focused vs. expansionary, civil vs. uncivil).} However, no prior work has used structural patterns to measure or predict members' community-level attitudes, such as SOVC.

Drawing on this related work, this paper fills a key research gap by modeling the relationship between community-level perceptions of SOVC and conversation-level markers of structure and style.

\section{Methods}\label{sec:methods}
We conducted our study on Reddit, a large, global platform hosting over 100,000 active communities around various topics and interests, independently created and managed by users called moderators. The platform's threaded, comment-driven discussion structure and voting features enable rich conversational interactions. Combined with the platform's diversity of communities, this makes Reddit an ideal context for studying SOVC.

\begin{table}[t]
\centering
\small
\renewcommand{\arraystretch}{1.2} % Adjust row height for better readability
\setlength{\tabcolsep}{10pt} % Adjust horizontal padding for columns
\begin{tabular}{|l|c|c|}
\hline
\textbf{Weekly Subscribed Visitors} & \textbf{Everyone} & \textbf{Mature} \\ \hline
1-800                               & 22.3\%            & 3.3\%           \\ \hline
800-2000                            & 22.7\%            & 3.7\%           \\ \hline
2000-6400                           & 21.0\%            & 3.3\%           \\ \hline
6400+                               & 20.3\%            & 3.3\%           \\ \hline
\end{tabular}
\caption{The 300 target subreddits were selected to be broadly representative of the broader target subreddit population, based on segments by size and rating (Everyone / Mature).}
\label{tab:target-subs}
\end{table}

\subsection{Measuring Sense of Virtual Community}\label{sec:methods_survey}
\subsubsection{Survey Design and Participants.}
We measured SOVC using a survey distributed via Qualtrics in 2023 to 2,826 Reddit users across 281 subreddits. Participants were selected randomly from subscribers to a sample of 300 public, English-language, safe-for-work (not containing sexually-explicit or gore content) subreddits, stratified based on activity and subscriber count. Eligible participants were over 18, located in the US, Canada, Australia, or the UK, with accounts in good standing and eligible to be contacted for a survey. Respondents were compensated with \$10 gift cards.

The survey asked about (1) informed consent, (2) demographics, and (3) community-specific questions. The full survey instrument is included in Appendix~\ref{app:instrument} and summarized here.
Users self-reporting that they were under 18 years old were skipped to the end and excluded from participation. The community-specific measures included 14 SOVC items, introduced by \citet{perkins1990sci} and extensively evaluated in prior work \cite{obst2002scifi1, obst2004revisiting, blanchard2007developing, abfalter2012sovc, kairam2022social}). Participants responded to each using a 5-point Likert scale (-2 = `Strongly disagree' to +2 = `Strongly agree').

Respondents self-reported their age using predefined ranges; the median age was `25-34', with 46\% (1,305) reporting that they were 35 or older. 91\% opted to self-disclose their gender identity; of these, 65\% identified as men, 21\% as women, and 4\% as non-binary (the remainder choosing to self-describe). These proportions of gender identity fairly closely resemble those of Reddit more broadly \cite{pew_2016_reddit_demographics}.
89\% self-reported that they had been active on the platform for at least a year; just 3\% reported a tenure of 3 months or less. 81\% indicated that they visited Reddit at least once per day. Just 4\% reported that they visited Reddit less often than weekly.

\subsubsection{Exploratory Factor Analysis.}
We removed 112 respondents (4.0\%) who `straight-lined' through the 14 SOVC measures,\footnote{\new{We removed all responses where the respondent provided the same answer to every single survey question.}} yielding 2,714 responses. \new{We analyzed responses using polychoric correlations to account for ordinal data (\textit{cf.} ~\cite{baglin2014efa}).} A Kaiser-Meyer-Olkin test ($MSA = 0.92$) and Bartlett's test ($p < 0.0001$) confirmed suitability for Exploratory Factor Analysis (EFA). Given the close expected relationship among factors, we used \texttt{promax} rotation, yielding a three-factor solution that explained 44\% of the variance in responses. McDonald's Omega indicates that a substantial portion of this variance (73\%) is explained by a common underlying dimension, supporting our expectation that these three distinct factors align with a shared construct of SOVC~\cite{kalkbrenner2023alpha}.

\begin{figure}[t]
    \centering
    %  trim={<left> <lower> <right> <upper>}
    \includegraphics[width=\linewidth,trim={0 8in 4.37in 0},clip]{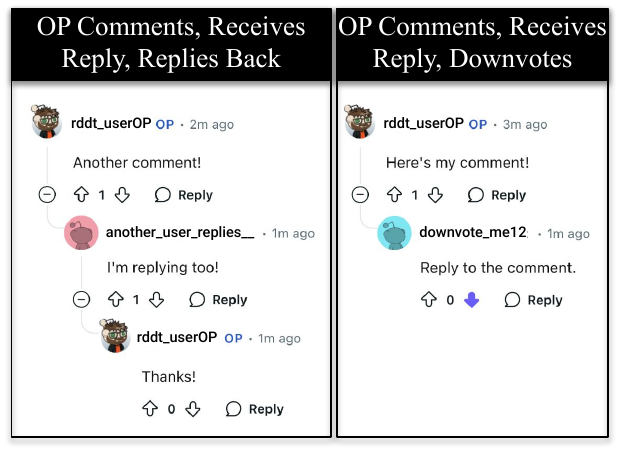}
    \caption{Conversational patterns include different interactions between community members. The left example here shows OP commenting, receiving a reply, and replying back, while the right example shows OP commenting, receiving a reply, and downvoting (blue arrow). A complete list of interactions is given in Appendix~\ref{app:patterns}.}
    \label{fig:patterns_example}
\end{figure}

\subsection{Capturing Online Behavioral Activity}\label{sec:methods_features}
To explore relationships between community interactions and SOVC, we collected a range of individual- and community-level behavioral trace data over a 30-day period preceding the survey. These data were used to construct our features capturing conversational structure and linguistic style, along with control variables to account for large-scale differences in community participation and size.

\subsubsection{Conversational Patterns.}
To capture the dynamics of user interactions within communities, we computed conversational patterns derived from chains of interactions initiated by posts or comments in the target subreddit. Following prior work on interaction motifs (\eg \cite{coletto2017automatic}), we identified recurring patterns of engagement that reflect different types of participation and response behaviors. For each post or comment by a user, we tracked subsequent interactions to form conversational patterns.
Using this approach, we identified eight distinct conversational patterns, representing variations in interaction dynamics (examples in \fig\ref{fig:patterns_example}, for a full list of patterns, see Appendix~\ref{app:patterns}), including differentiation between patterns that start with a post (a top-level submission on Reddit) and those that start with a comment (a comment that is a reply to a post or another comment). \new{For each subreddit, we calculated the total occurrences of these patterns over a 30-day observation period occurring immediately prior to the SOVC survey.} Conversational pattern counts were normalized by the total number of interactions within each subreddit, yielding an 8-element proportion vector that ensures comparability across communities with varying activity levels.

% 2,342,651 total posts and comments analyzed for patterns
\new{
We selected this method for computing conversational patterns as it is both established in prior work \cite{coletto2017automatic} and feasible to implement at a very large scale. Our analyses included 2,342,651 posts and comments across our 281 participating subreddits. While such methods have been used to predict conversational level outcomes \cite{coletto2017automatic}, ours is the first work that we are aware of to use conversational patterns to predict outcomes at the community level.
}

\subsubsection{Linguistic Style.}
Linguistic style was analyzed using LIWC-22, a validated tool for computational text analysis that derives stylistic and psychological features from language~\cite{liwc22}. We apply LIWC-22 to all posts and comments in each community from the target 30-day period, mapping words to one or more of 118 different categories, such as pronouns, emotion words, and perception-related words. \new{Following precedent and guidance from prior work (\eg \citet{khalid2020style, matthews2015they}), we include all LIWC features except 19 topic-specific LIWC categories (the culture, lifestyle, and physical dictionaries from the LIWC extended dictionary) to avoid topical leakage and focus on stylistic features,} which are more generalizable across domains and strongly linked to social phenomena.

\new{
LIWC was selected for quantifying linguistic style as it has been extensively used and validated for a broad range of applications \cite{liwc22} as well as being suitable to apply to text at a massive scale. Our analysis of linguistic style included more than 70 million tokens, a scale which renders many alternative methods computationally intractable.  While methods such as  word embeddings can be used to compare language between different users or communities \cite{Waller2020QuantifyingSO, danescu2013country}, in this work we use an absolute measurement of language at the community level to predict SOVC, an important community outcome.
}

\subsubsection{Control Variables.}
At the individual level, we captured measures of activity, \new{both across all of Reddit (\eg the number of unique communities visited, the frequency of contribution) -- and within the target community (\eg duration of membership tenure, whether or not they had recently voted).} At the community level, we collected aggregate measures including the total number of visitors, contributors, and daily active users. These control features were included in our models to adjust for baseline differences across users and subreddits.

\subsection{Modeling Approach}\label{sec:methods_modeling}
\new{To predict our measures of SOVC, we adopted a hierarchical linear model approach with a random slope} and intercept, integrating individual-level and community-level features while accommodating the nested data structure. We applied LASSO (Least Absolute Shrinkage and Selection Operator) for feature selection, which identifies the most predictive behavioral features by setting irrelevant coefficients to zero. We used 5-fold cross-validation to select the best alpha penalty for the L1 norm of the coefficients. \new{Features were standardized (zero mean, unit variance) to facilitate interpretation of coefficients and ensure comparability across predictors.}

The model itself was constructed with two levels:
\begin{itemize}
    \item \textbf{Individual-Level.} A mixed-effects model used individual-level features (\eg, voting behavior, communities visited) to predict SOVC scores, with random intercepts accounting for subreddit-specific variability.
    \item \textbf{Community-Level.} A LASSO-regularized linear model used community-level features (\eg, community age, number of visitors) to predict the subreddit random intercepts obtained from the first model.
\end{itemize}

\new{
We selected a hierarchical linear model due to the natural bilevel nature of our features, which describe both individuals (respondents to our SOVC survey) and communities, each of which had several respondents.
}

\new{
More formally, the individual level SOVC score $\hat{y}_{ij}$ is given by:
\[
\hat{y}_{ij} = \beta_1x_{ij1} + \beta_2x_{ij2} + \cdots + \beta_nx_{ijn} + \alpha_i
\]
}
\noindent
\new{
The community-level random intercept $\hat{\alpha}_i$ is given by:
\[
\hat{\alpha}_i = \gamma_1x_{i1} + \gamma_2x_{i2} + \cdots + \gamma_mx_{im}
\]
}
\noindent
\new{
Where $x_{ijn}$ represents that $n$-th value of community $i$'s member $j$'s feature vector, $x_{im}$ represents the $m$-th value of community $i$'s feature vector, and $\beta_n$ and $\gamma_m$ represent the learned linear coefficients selected using LASSO-regularization.
}

Model performance was assessed using adjusted $R^2$ and variance partitioning to evaluate the relative importance of the individual- and community-level predictors. An ablation study quantified the relative importance of conversational structure and linguistic style features, confirming that both contributed significantly to SOVC predictions.

\subsection{Ethical and Privacy Considerations}\label{sec:ethics}
Our study followed strict ethical guidelines to protect participant privacy and ensure responsible data use, as confirmed by an internal review of our study procedures at Reddit, Inc.
Surveys were targeted to users whose selected settings made them eligible to be contacted, and survey participants provided informed consent. All data were collected and processed under Reddit's privacy policy. 
Behavioral trace data were anonymized and aggregated to prevent reidentification and analysis focused exclusively on either high-level individual-level features (\eg vote counts \vs individual votes) or aggregated community-level patterns (\eg LIWC category counts vs. individual text strings). We believe the potential for negative social impacts and misuse from our work is very limited.

\section{Results}\label{sec:results}

% putting this here for layout -g
\begin{table*}[t]
\centering
\begin{tabular}{|p{12cm}|c|c|c|}
\hline
\textbf{Scale} / Item &
  \multicolumn{1}{c|}{\begin{tabular}[c]{@{}c@{}}Factor 1\end{tabular}} &
  \multicolumn{1}{c|}{\begin{tabular}[c]{@{}c@{}}Factor 2\end{tabular}} &
  \multicolumn{1}{c|}{\begin{tabular}[c]{@{}c@{}}Factor 3\end{tabular}} \\ \hline
\multicolumn{4}{|l|}{\textbf{MB: Membership \& Belonging ($\alpha = 0.698$)}} \\ \hline
I expect to be a part of this community for a long time. & \textbf{0.79} & -0.04 & -0.04 \\
I think this community is a good thing for me to be a part of. & \textbf{0.64} & 0.19 & -0.09 \\
It is important to me to be a part of this community. & \textbf{0.49} & 0.01 & 0.34 \\
I feel at home in this community. & \textbf{0.44} & 0.18 & 0.14 \\ \hline
\multicolumn{4}{|l|}{\textbf{CSV: Cooperation \& Shared Values ($\alpha = 0.795$)}} \\ \hline
If there is a problem in this community, members can get it solved. & -0.07 & \textbf{0.70} & -0.01 \\
Members of this community can be counted on to help others. & 0.01 & \textbf{0.70} & -0.04 \\
I want the same things from this community as other members. & 0.21 & \textbf{0.43} & -0.11 \\
Members of this community share the same values. & 0.07 & \textbf{0.50} & 0.00 \\ \hline
\multicolumn{4}{|l|}{\textbf{CI: Connection \& Influence ($\alpha = 0.792$)}} \\ \hline
Most members of this community know me. & -0.07 & -0.20 & \textbf{0.88} \\
I have friends in this community that I can depend on. & -0.03 & 0.04 & \textbf{0.73} \\
I recognize the screen names of most participants in this community. & 0.04 & -0.07 & \textbf{0.67} \\
I feel like I have influence over what this community is like. & 0.00 & 0.12 & \textbf{0.53} \\
I care about what other community members think of me. & 0.13 & 0.03 & \textbf{0.45} \\
If I have a personal problem, I can turn to members of this community. & -0.12 & 0.38 & \textbf{0.41} \\ \hline
\end{tabular}
\caption{ Dimensions of SOVC across subreddits as identified through EFA, with factor loadings for items and Cronbach's $\alpha$ for factors. All items were tested using a 5-point Likert scale ranging from ``-2: Strongly Disagree'' to ``+2: Strongly Agree''.}
\label{tab:factor-loadings}
\end{table*}

\subsection{Characterizing SOVC on Reddit}\label{sec:results_sovc_factors}
\new{To characterize SOVC on Reddit, we performed an Exploratory Factor Analysis (EFA, described in \sect\ref{sec:methods_survey}) on our survey data to} identify three dimensions of Sense of Virtual Community (SOVC) on Reddit, which we term \textbf{Membership \& Belonging} (MB), \textbf{Cooperation \& Shared Values} (CSV), and \textbf{Connection \& Influence} (CI). These dimensions, summarized in Table~\ref{tab:factor-loadings}, align with established constructs in the literature and capture distinct, complementary aspects of community experience.

% putting here for layout
\begin{figure}[t]
    \centering
    \includegraphics[width=\linewidth]{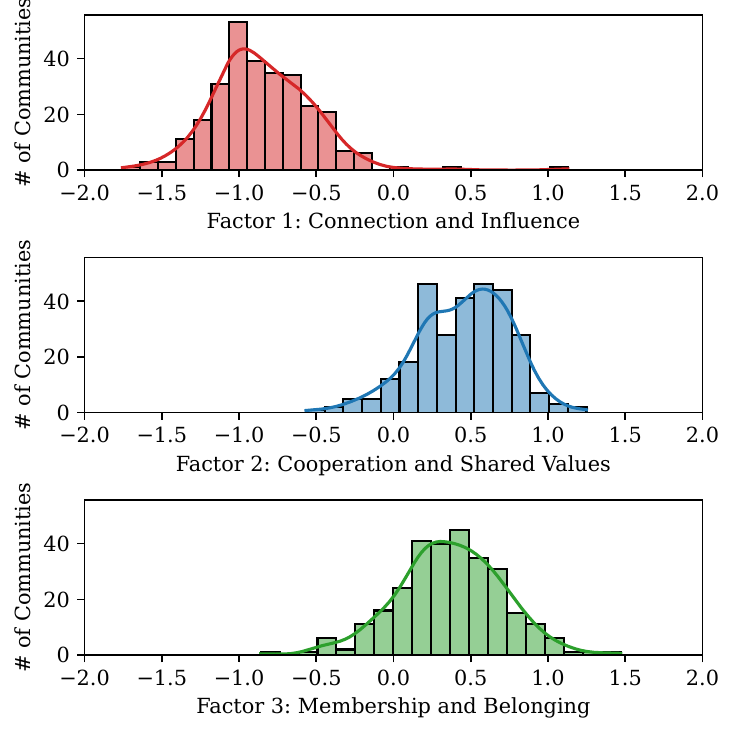}
    \caption{ Community members generally rate their connection and influence (Factor 1) as lower than their cooperation and shared values (Factor 2) or membership and belonging (Factor 3). This figure shows the distribution of communities' SOVC scores for each factor, computed by averaging the responses for each community.}
    \label{fig:factor_distribution}
\end{figure}

\textbf{Membership \& Belonging}  \new{reflects the extent to which individuals feel embedded within a community. Items loading on this factor have previously been categorized as `belonging'~\cite{obst2002scifi1, kairam2022social}, `membership'~\cite{obst2004revisiting}, `conscious identification'~\cite{obst2002scifi1}, and `shared emotional connection'~\cite{abfalter2012sovc}.}

\textbf{Cooperation \& Shared Values}  \new{captures perceptions of collective alignment and mutual support. Items loading on this factor have been categorized in prior work under various constructs, including `cooperation/shared values', emotional connection, `influence',  `integration/needs fulfillment', and `cohesion'.}

\textbf{Connection \& Influence}  \new{relates to interpersonal relationships and perceived social influence within the community. Items in this factor have been categorized in prior work as a variety of constructs related to SOVC, including `leadership/influence', `friendship \& social support', and `integration/needs fulfillment'. }

As shown in Table~\ref{tab:factor-loadings}, the inter-item reliability for these scales is moderate ($\alpha_{MB} = 0.698$; $\alpha_{CSV} = 0.795$; $\alpha_{CI} = 0.792$). While $\alpha_{MB}$ can be improved substantially by dropping the least-correlated item \textit{`I feel at home in this community'}, we have retained this item given its strong alignment with this construct in prior work.

% put here for layout
\begin{table}[t]
    \centering
    {\footnotesize
    \begin{tabular}{lcccc}
    \hline
    \textbf{Model}       & \textbf{CI} & \textbf{CSV} & \textbf{MB} & \textbf{Overall} \\ \hline
    Control Only         & 0.152                 & 0.095                  & 0.054                 & 0.100                     \\ 
    + Style              & 0.311                 & 0.326                  & 0.272                 & 0.303                     \\ 
    + Structure          & 0.171                 & 0.197                  & 0.113                 & 0.160                     \\ 
    + Style \& Structure                 & 0.341                 & 0.348                  & 0.305                 & 0.331                     \\ \hline
    \end{tabular}}
    \caption{ {Ablation Study Results.} \(R^2\) values capturing explained variance for each SOVC dimension (CI: Connection \& Influence, CSV: Cooperation \& Shared Values, MB: Membership \& Belonging) and overall, for models with different feature combinations. The ``Control Only'' model includes no structural or stylistic features. Stylistic features contribute more to \(R^2\) than structural features alone, \new{with the full model (Control + Style \& Structure) achieving the highest overall performance.}}
    \label{tab:ablation_revised}
\end{table}

\subsection{Quantifying SOVC on Reddit}\label{sec:results_survey}
\new{We quantify SOVC for modeling by mapping responses from -2 (Strongly Disagree) to +2 (Strongly Agree) and averaging scores across items in a scale. This lets us compute for each respondent a score for each factor.} These scores represent the dependent variables in our individual-level and community-level SOVC models.
As shown in \fig\ref{fig:factor_distribution}, all three factors follow a roughly normal distribution.

SOVC scores varied significantly across subreddits and respondents (Figure~\ref{fig:factor_distribution}). Communities scored highest on \textbf{Cooperation \& Shared Values} ($\mu = 0.50, \sigma = 0.61$) and \textbf{Membership \& Belonging} ($\mu = 0.35, \sigma = 0.72$), with relatively lower scores on \textbf{Connection \& Influence} ($\mu = -0.81, \sigma = 0.70$). These results suggest that personal connections and influence among members may be less prominent than generalized feelings of belonging or cooperation in the communities studied.

% putting this here for layout -g
\begin{table*}[t]
\centering
\renewcommand{\arraystretch}{1.2}
\setlength{\tabcolsep}{5pt}
\begin{tabular}{|l|l|c|c|c|c|}
\hline
\textbf{Model Level} & \textbf{Feature} & \textbf{CI (\(\beta, p\))} & \textbf{CSV (\(\beta, p\))} & \textbf{MB (\(\beta, p\))} \\ \hline
\multirow{8}{*}{\textbf{Individual Level}} 
 & \multicolumn{4}{|l|}{\textit{Within-Community Features}} \\ \cline{2-5}
 & Days Active (past 30)                    & \(0.005**\)    & \(0.001\)   & \(0.006**\)    \\
 & Length of Membership (\(\log_2\))        & \(-0.001\)    & \(0.006\)   & \(0.017**\)    \\
 & Has Voted (T/F)                          & \(0.049\)     & \(0.072*\)   & \(0.077*\)     \\
 & Has Posted or Commented (T/F)            & \(0.108**\)    & \(0.023\)   & \(0.092*\)     \\ \cline{2-5}
 & \multicolumn{4}{|l|}{\textit{Sitewide Features}} \\ \cline{2-5}
 & Communities Visited (\(\log_2\))         & \(-0.030**\)   & \(-0.017\)  & \(-0.029**\)   \\
 & Number of Votes (\(\log_2\))             & \(-0.023***\)  & \(-0.005\)  & \(-0.004\)    \\
 & Number of Posts and Comments (\(\log_2\))& \(0.021**\)    & \(-0.013*\)  & \(-0.006\)    \\ \hline
\multirow{7}{*}{\textbf{Community Level}} 
 & \multicolumn{4}{|l|}{\textit{Control Features}} \\ \cline{2-5}
 & Community Age (days)                    & \(-0.121*\)    & \(-0.050\)  & \(-\)         \\
 & Visitors (past 30 days, \(\log_2\))      & \(-0.185*\)    & \(-\)       & \(-\)         \\ \cline{2-5}
 & \multicolumn{4}{|l|}{\textit{Conversational Structure Features}} \\ \cline{2-5}
 & OP Posts, Receives Reply, Replies Back    & \(-\)        & \(0.182*\)   & \(0.090\)     \\
 & OP Comments, Receives Reply, Replies Back & \(0.234**\)   & \(-\)       & \(-\)         \\
 & OP Comments, Receives Reply, Upvotes      & \(-\)        & \(-\)       & \(0.141*\)     \\ \cline{2-5}
 & \multicolumn{4}{|l|}{\textit{Linguistic Style Features (LIWC)}} \\ \cline{2-5}
 & Third Person Plural Pronouns (they)           & \(0.127*\)    & \(0.122*\)   & \(0.118\)     \\
 & Prosocial Behavior (care, thank, help)        & \(0.166*\)    & \(0.096\)   & \(0.124\)     \\
 & Female References (she, girl, woman)          & \(-0.171*\)   & \(-0.204*\)  & \(-0.284***\)  \\
 & Lacks (don’t have, didn’t have, less)         & \(-0.127*\)   & \(-\)       & \(-0.093\)    \\
 & Feeling (feel, hard, cool, felt)              & \(0.211**\)   & \(0.138*\)   & \(0.099\)     \\
 & Future Focused (will, going to, have to)      & \(-\)        & \(0.016\)   & \(0.187**\)    \\
 & Netspeak (u, lol, haha, emoji)                & \(-\)        & \(0.156*\)   & \(0.167*\)     \\ \hline
\end{tabular}

\caption{ $\beta$ coefficients for models predicting Connection \& Influence (CI), Cooperation \& Shared Values (CSV), and Membership \& Belonging (MB) scores, based on structure \& style features, along with individual- and community-level controls. Features are included in the model only if significantly associated, on their own, with the outcome variable. Many coefficients are shrunk to zero via Lasso regularization. Indicators for $p$-values are as follows: * : $p < 0.05$,
** : $p < 0.01$, *** : $p < 0.001$.}
\label{tab:model-coeffs}
\end{table*}

\new{
While studying SOVC independently from the topic of a community is the primary goal of this work, it's important to understand how SOVC varies across communities of different topics. To do so, we use Reddit's publicly available topics. More details on topic classification and extended results are included in Appendix~\ref{app:topic}.
}

\new{
Subreddits in our sample which scored highest on Connection \& Influence include Business \& Finance Communities and Nature \& Outdoors communities. Health communities and those focused on Vehicles had the highest sense of Cooperation \& Shared Values, while Health and Nature \& Outdoors communities had some of the strongest senses of Membership \& Belonging (Appendix~\ref{app:topic}). However, the primary focus of this work is on how SOVC varies across communities based on their conversational structure and linguistic style. Future work could specifically examine SOVC between communities with different topics (\sect\ref{sec:limitations}).
}

\subsection{Modeling SOVC with On-Platform Behavior}\label{sec:results_modeling}
Using our hierarchical linear modeling approach, \new{we predicted each of the three SOVC dimensions, as measured by the EFA factor scores computed above, using individual- and community-level measures of conversation structure and linguistic style (\sect\ref{sec:methods_features}). We find that both conversational structure and linguistic style separately capture more SOVC than control features only, and that linguistic style features contribute more to model performance than conversational structure features (Table~\ref{tab:ablation_revised}). Linguistic style features on their own capture more than 30\% of the variance in SOVC.}

\begin{figure*}[t]
    \centering
    \includegraphics[width=.9\textwidth]{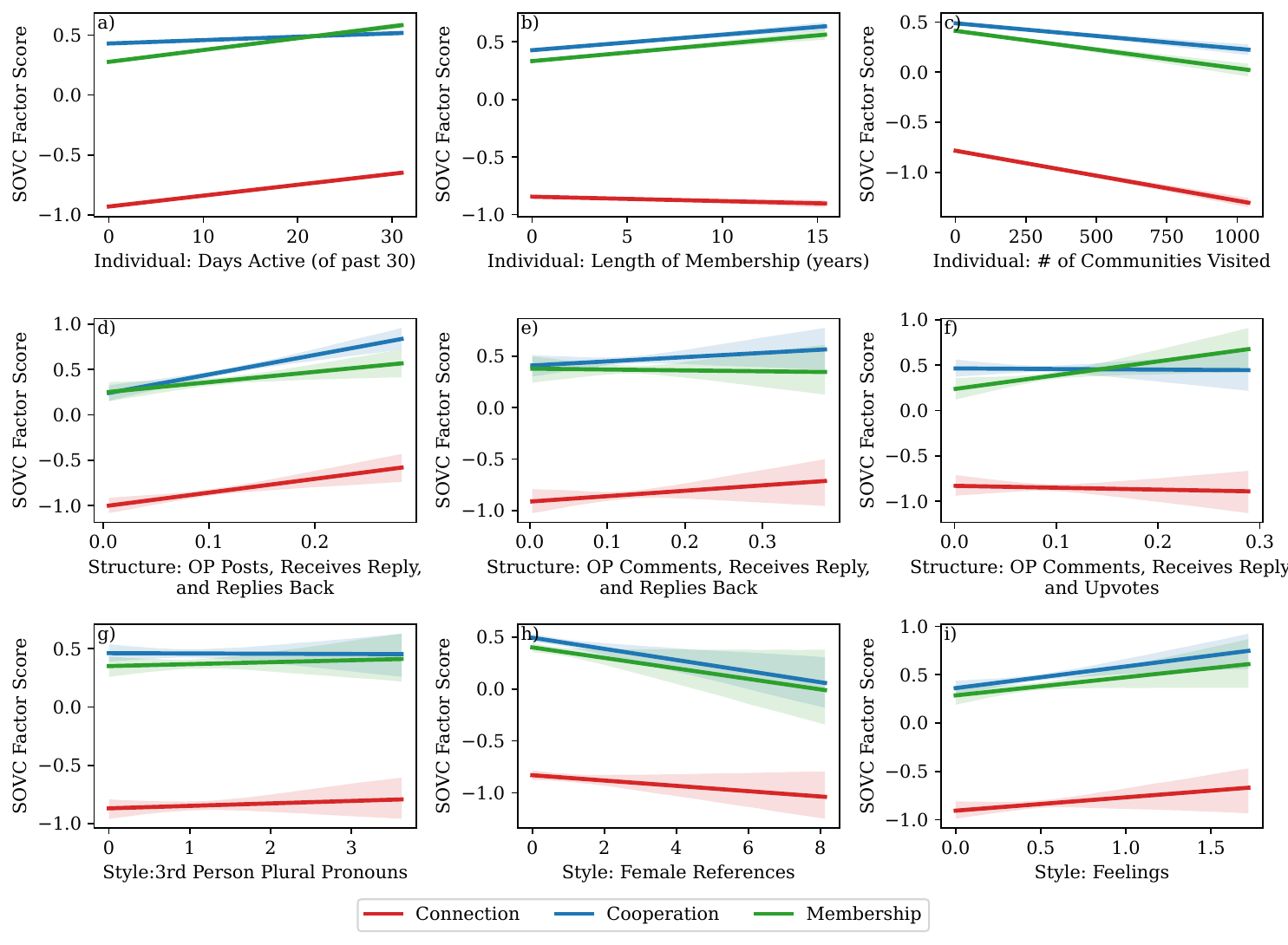}
    \caption{ Relationships between selected individual and community level features and model-predicted SOVC. The top row shows individual-level features, while the bottom two rows show community-level conversational structure and linguistic style, respectively. \new{Conversational structure axes show the proportion of all patterns that match the pattern on the X axis label. Linguistic Style axes units are taken directly from LIWC output.}}
    \label{fig:model_results_combined}
\end{figure*}

\new{
Our full SOVC model, including both conversational structure and linguistic style features, captures 33.1\% of the variance in SOVC across communities and individuals. As communities vary widely in their topic, membership, and values \cite{weld2021better}, we believe that it is noteworthy that our 19 readily quantifiable features capture nearly one-third of variance in SOVC.
}

The coefficients for our linear model, shown in Table~\ref{tab:model-coeffs}, show the strength of the relationship between each measure of community activity and self-reported SOVC. We illustrate the relationship between features discussed below and model predictions for SOVC scores in Figure~\ref{fig:model_results_combined}.

\new{
Among individual-level features, \textbf{days active} in the community positively predicted both Connection \& Influence and Membership \& Belonging, while longer \textbf{membership tenure} was uniquely associated with higher Membership \& Belonging.} \textbf{Active participation}, such as posting/commenting, was positively associated with all three SOVC dimensions, highlighting its centrality to community experience. Interestingly, greater \textbf{cross-subreddit participation} tended to be associated with lower Cooperation \& Shared Values, suggesting that cross-community engagement may dilute perceptions of alignment within a single subreddit.

\subsubsection{Conversational Structure.}
Our analysis reveals nuanced relationships between conversational structures and the dimensions of Sense of Virtual Community (SOVC). Notably, communities with more reciprocal reply chains receive higher scores for Cooperation \& Shared Values and Membership \& Belonging (when they start on posts) and Connection \& Influence (when they start on comments). This result emphasizes the role of sustained dialogue in fostering interpersonal relationships and collective alignment.
Increased reciprocal conversation has been linked to various positive community outcomes, including discussion quality~\cite{fisher2006you, janssen2005online} and provision of support~\cite{althoff2014ask}.

\new{
On the other hand, we find relatively limited relationships between voting behavior and SOVC. We do not find that downvoting behavior is significantly associated with any dimension of SOVC, while more frequent upvotes to replies are only associated with one of the three dimensions of SOVC; increased Membership \& Belonging.
Upvotes on Reddit serve two functions which may impact membership/belonging: (1) providing positive feedback to the upvoted commenter that enhances their pro-community attitudes \cite{davis2021emotional, lambert_2025_positive_reinforcement}, and (2) aligning community expectations by making desirable content more visible \cite{goyal2024uncovering}.}

\subsubsection{Linguistic Style.}
\new{
Features capturing community-wide differences in linguistic style were highly informative, explaining more than 30\% of the variance in self-reported SOVC scores.}
Stronger feelings of Connection \& Influence are associated with the use of more third-person plural pronouns (\eg \textit{they}).
Prosocial language (\eg, `thank', `help') is positively associated with Connection \& Influence and Cooperation \& Shared Values, aligning with prior work suggesting that gratitude language in online communities can create positive feedback loops, prompting future prosocial behavior~\cite{makri2020can,althoff2014ask}. However, more references to wants or lack of things (\eg \textit{don't have}, \textit{less}) are associated with \textit{lower} Connection \& Influence and Membership \& Belonging.

Expressions of sensation or emotion (\eg, `feel', `hard') align positively with Connection \& Influence and Cooperation \& Shared Values. Positive expressions of emotion can certainly indicate effective functioning as a collective~\cite{fischer2007linguistic, krifka2003group}; this may also reflect how strong, anonymous communities empower members to share negative emotional experiences to obtain support~\cite{de2014mental, ammari2019self}. 

We note a positive relationship between Netspeak (\eg \textit{u}, \textit{lol}, emoji) and `future-focused' language and feelings of Membership \& Belonging. Prior distinctions between `common identity' and `common bond' groups define the former as focused more on the goal and purpose of the group than connections to other members~\cite{sassenberg2002common}. \new{
Fewer references to `girls' or `women' corresponded to stronger SOVC across all three dimensions; this may reflect that communities which adopt more inclusive language may tend to function better overall, although additional research is needed to understand the role that gendered language plays in community members' SOVC.
}

In \fig~\ref{fig:model_results_combined}, we visualize the relationships between selected individual and community level features and the three factors of SOVC. \fig~\ref{fig:model_results_combined}b shows how community members with longer lengths of membership have higher senses of cooperation and membership, but lower senses of connection.  \fig~\ref{fig:model_results_combined}d,e,f shows how greater conversational depth and more upvotes by OP are associated with higher SOVC, especially connection and cooperation.
 \fig~\ref{fig:model_results_combined}h shows how more female references are associated with lower SOVC, while more discussion of feelings are associated with higher SOVC across all three factors (\fig~\ref{fig:model_results_combined}i).

\section{Discussion}\label{sec:discussion}
In this work, we sought to understand how \textit{Sense of Virtual Community} (SOVC) varies across online communities, and identify associated structural and stylistic markers of interactions. Using self-reported community perceptions from 2,826 active Reddit users, regarding 281 unique communities in which they participated, our hierarchical model allowed us to identify how differences in conversational structure and linguistic style can predict community-level variation in SOVC. Below, we present a brief summary of our results and discuss some implications for researchers, designers, and practitioners who share an interest in fostering more positive social experiences online.

\subsection{Summary of Results}

\subsubsection{Dimensions of SOVC on Reddit.}
Through exploratory factor analysis, we identified three dimensions of SOVC -- \textbf{Membership \& Belonging} (MB), \textbf{Cooperation \& Shared Values} (CSV), and \textbf{Connection \& Influence} (CI) -- that align well with established constructs from prior work. Membership \& Belonging captures the extent to which individuals feel integrated within their community, aligning with notions of belonging and shared emotional connection.
Cooperation \& Shared Values reflects collective alignment and mutual support, 
Connection \& Influence represents interpersonal relationships and the perceived ability to impact the community.
Researchers should consider how these different dimensions of SOVC may be in tension with one another, requiring trade-offs to be made when working to improve communities \cite{weld2024taxonomy}.

\subsubsection{Distribution of SOVC Scores.}
SOVC scores varied substantially across communities and respondents. The higher scores observed for Cooperation \& Shared Values and Membership \& Belonging than for Connection \& Influence suggest that community experience on Reddit has more to do with fostering shared goals and a broader sense of belonging than prompting strong interpersonal connections or influence as an individual on the community. 
\new{We note our strategy for sampling communities in this work necessarily means that the findings are not broadly representative across all Reddit communities. In particular, we don't include small communities in which interpersonal connection may be more likely to occur \cite{sheng2020twitch}. Future work could look explicitly at small communities.}

\subsubsection{Modeling SOVC with Linguistic Style and Conversational Structure.}
% ~3 takeaways:
Our hierarchical model is able to predict 33.1\% of variance in SOVC using automatically quantifiable features. While both linguistic style and conversational structure features are important to our model's predictive performance, linguistic style plays a larger role in modeling SOVC. \new{Future work using LLM-based measurement of features may capture even more variance in SOVC.}
Regarding conversational structure, deeper conversational depth is more strongly associated with stronger SOVC than voting behavior. We found downvoting behavior is not significantly associated with any dimension of SOVC, while upvoting behavior is only associated with Membership \& Belonging. Regarding linguistic style, more third person pronouns (they) and more discussion of prosocial behavior are especially strongly associated with increased SOVC, being positively associated with all three dimensions.

\subsection{Practical Implications}
Our findings offer actionable insights for community moderators and platform designers to foster stronger communities.

\subsubsection{Implications for Community Moderators.} 
Given the outsized role that moderators play in shaping community experiences on Reddit and other platforms through community-building activities~\cite{kairam2024founder, seering2023moderates}, these findings suggest certain actions that moderators could take within their communities to foster behaviors associated with higher SOVC. Moderators might foster more reciprocal reply chains by initiating and promoting dialogues through Q\&A threads (\eg AMAs, `ask me anything') or collaborative discussion formats. \new{Moderators can emphasize norms of appreciation and recognition, such as rewarding thoughtful responses with upvotes or featuring top contributors \cite{lambert_2025_positive_reinforcement}.} Engaging community members in shared projects, such as those supported by Reddit Community Funds, could engage community members in goal-oriented, future-tense discussion that build a sense of Membership \& Belonging. Moderators could easily support the use of more inclusive language through rules set up in Post Guidance~\cite{ribeiro2024post} or similar proactive moderation tools on other platforms.
\new{Moderators are also uniquely familiar with their community's specific attributes and needs \cite{weld2021better, cullen2022moderation}; future work could explore how moderators can support SOVC in community-specific ways.}

\subsubsection{Design of Community Features.} \new{These findings also inform the design of product features that can increase reciprocal interaction and positive engagement. On-platform prompts or off-platform notifications encouraging community members to reply to messages, especially those from first-time posters, might help to increase SOVC for both the posters and the repliers. Automated summaries or highlights from active threads could be used to help direct users to opportunities to reply and foster reciprocal engagement. Designers could encourage positive feedback by explicitly rewarding users who promote belonging by upvoting others' contributions through badges or flair. Hiding content that has received a certain number of downvotes could prevent the accumulation of additional downvotes, potentially limiting further impacts on feelings of cooperation and membership.}

\subsubsection{Leveraging SOVC scores.} Finally, this work offers a content-agnostic way to compute SOVC scores per community on an ongoing basis, which could be leveraged for purposes outside of the communities themselves. These signals can be included in community recommendation algorithms to help drive new users towards communities that offer more potential to drive feelings of connection, cooperation, and belonging depending on user needs \& preferences. Tracking SOVC over time could offer insight into how community health is changing over time, for individual communities and in aggregate across the platform, enabling social platforms to intervene, when needed. SOVC scores could be used as metrics in experiments to evaluate feature rollouts to determine if new features are successfully driving behaviors associated with stronger community attitudes.

\subsection{Limitations and Future Work}\label{sec:limitations}
We acknowledge several limitations of our study, which may also offer opportunities for future research. Our analysis focused on a subset of Reddit communities selected based on activity and size, which may not represent the broader spectrum of smaller or less active subreddits; prior work has found that communities with features characteristic of `common-bond' groups may naturally limit their growth~\cite{kairam2012life}, which would leave them underrepresented in our sample. Future work should explore whether the identified patterns generalize to these contexts, \new{as well as other platforms with distinct community norms and structures, such as different voting affordances (\eg Likes in Facebook Groups \textit{vs.} upvotes on Reddit}). While our content-agnostic approach supports generalization across community topics, it remains limited to English-language data. Cross-cultural studies are needed to assess how linguistic and conversational patterns predict SOVC in non-English or multilingual communities, potentially revealing cultural nuances in community dynamics. \new{Our modeling approach identifies meaningful associations between community behavior and SOVC, but cannot establish causality; future experimental studies could clarify causal relationships.}
\new{
We deliberately use relatively simple methods, such as LIWC, to quantify conversational structure and linguistic style, as these methods are wide used, well validated, and computationally tractable given the large scale of our study. However, future work could incorporate additional and more sophisticated features, such as more complex conversational patterns, LLM-based measurements of linguistic style, and additional control features to measure community topic directly. Such methods may be able to predict a larger fraction of the variance in SOVC.
}
\section{Conclusion}\label{sec:conclusion}
This study advances the understanding of Sense of Virtual Community (SOVC) by providing the first large-scale analysis of how conversational structure and linguistic style shape community experiences across diverse Reddit communities. Using a content-agnostic approach, we identified three dimensions of SOVC—Membership \& Belonging, Cooperation \& Shared Values, and Connection \& Influence—and demonstrated that interaction patterns and linguistic features explain a substantial portion of their variance.

Our findings emphasize the importance of reciprocal interactions, prosocial language, and future-oriented communication in fostering stronger online communities. These insights deepen theoretical understanding of how online communities function and offer actionable strategies for moderators and platform designers to enhance community engagement, cohesion, and inclusivity.

In contrast to prior work focused on specific content or platforms, this study establishes generalizable principles that are applicable across a diverse set of communities and community members. By demonstrating how online interactions influence community attachment, we highlight the transformative potential of digital communities to promote individual and collective well-being. Future research should build on these findings to explore causal mechanisms and extend their applicability to broader %cultural and linguistic
contexts.

\section*{Acknowledgments}
This research was supported by the Office of Naval Research (\#N00014-21-1-2154), NSF grant IIS-1901386, NSF CAREER IIS-2142794, NSF grant CNS-2025022, and the Bill \& Melinda Gates Foundation (INV-004841).

{ \small
\bibliography{bibliography}
}
\section*{Paper Checklist}

\begin{enumerate}

\item For most authors...
\begin{enumerate}
    \item  Would answering this research question advance science without violating social contracts, such as violating privacy norms, perpetuating unfair profiling, exacerbating the socio-economic divide, or implying disrespect to societies or cultures?
    \answerYes{Yes, we make several important contributions (\sect\ref{sec:intro}) while taking great care to protect the privacy of our participants and be respectful of all cultures (\sect\ref{sec:ethics}).}
  \item Do your main claims in the abstract and introduction accurately reflect the paper's contributions and scope?
    \answerYes{Yes}
   \item Do you clarify how the proposed methodological approach is appropriate for the claims made? 
    \answerYes{Yes, discussed in \sect\ref{sec:methods} and \sect\ref{sec:limitations}.}
   \item Do you clarify what are possible artifacts in the data used, given population-specific distributions?
    \answerYes{Yes, discussed in \sect\ref{sec:methods_survey} and \sect\ref{sec:limitations}.}
  \item Did you describe the limitations of your work?
    \answerYes{Yes, described in \sect\ref{sec:limitations}.}
  \item Did you discuss any potential negative societal impacts of your work?
    \answerYes{Yes, discussed in \sect\ref{sec:ethics}.}
      \item Did you discuss any potential misuse of your work?
    \answerYes{Yes, discussed in \sect\ref{sec:ethics}.}
    \item Did you describe steps taken to prevent or mitigate potential negative outcomes of the research, such as data and model documentation, data anonymization, responsible release, access control, and the reproducibility of findings?
    \answerYes{Yes, discussed in \sect\ref{sec:ethics}.}
  \item Have you read the ethics review guidelines and ensured that your paper conforms to them?
    \answerYes{Yes.}
\end{enumerate}

\item Additionally, if your study involves hypotheses testing...
\begin{enumerate}
  \item Did you clearly state the assumptions underlying all theoretical results?
    \answerNA{Not Applicable.}
  \item Have you provided justifications for all theoretical results?
    \answerNA{Not Applicable.}
  \item Did you discuss competing hypotheses or theories that might challenge or complement your theoretical results?
    \answerYes{Yes, discussed in \sect\ref{sec:related} and \sect\ref{sec:limitations}.}
  \item Have you considered alternative mechanisms or explanations that might account for the same outcomes observed in your study?
    \answerYes{Yes, discussed in \sect\ref{sec:limitations}.}
  \item Did you address potential biases or limitations in your theoretical framework?
    \answerYes{Yes, discussed in \sect\ref{sec:limitations}.}
  \item Have you related your theoretical results to the existing literature in social science?
    \answerYes{Yes, discussed in \sect\ref{sec:related}.}
  \item Did you discuss the implications of your theoretical results for policy, practice, or further research in the social science domain?
    \answerYes{Yes, discussed in \sect\ref{sec:discussion}.}
\end{enumerate}

\item Additionally, if you are including theoretical proofs...
\begin{enumerate}
  \item Did you state the full set of assumptions of all theoretical results?
    \answerNA{Not Applicable.}
	\item Did you include complete proofs of all theoretical results?
    \answerNA{Not Applicable.}
\end{enumerate}

\item Additionally, if you ran machine learning experiments...
\begin{enumerate}
  \item Did you include the code, data, and instructions needed to reproduce the main experimental results (either in the supplemental material or as a URL)?
    \answerNA{Not Applicable.}
  \item Did you specify all the training details (e.g., data splits, hyperparameters, how they were chosen)?
    \answerNA{Not Applicable.}
     \item Did you report error bars (e.g., with respect to the random seed after running experiments multiple times)?
    \answerNA{Not Applicable.}
	\item Did you include the total amount of compute and the type of resources used (e.g., type of GPUs, internal cluster, or cloud provider)?
    \answerNA{Not Applicable.}
     \item Do you justify how the proposed evaluation is sufficient and appropriate to the claims made? 
    \answerNA{Not Applicable.}
     \item Do you discuss what is ``the cost'' of misclassification and fault (in)tolerance?
    \answerNA{Not Applicable.}
  
\end{enumerate}

\item Additionally, if you are using existing assets (e.g., code, data, models) or curating/releasing new assets, \textbf{without compromising anonymity}...
\begin{enumerate}
  \item If your work uses existing assets, did you cite the creators?
    \answerNA{We do not use existing assets.}
  \item Did you mention the license of the assets?
    \answerNA{We do not use existing assets.}
  \item Did you include any new assets in the supplemental material or as a URL?
    \answerNo{We do publish any new assets.}
  \item Did you discuss whether and how consent was obtained from people whose data you're using/curating?
    \answerYes{We discuss how consent as obtained in \sect\ref{sec:methods_survey} and \sect\ref{app:instrument}.}
  \item Did you discuss whether the data you are using/curating contains personally identifiable information or offensive content?
    \answerYes{Even thought we are not releasing new data, we discuss PII in \sect\ref{sec:ethics}.}
\item If you are curating or releasing new datasets, did you discuss how you intend to make your datasets FAIR?
\answerNA{We are not releasing new datasets.}
\item If you are curating or releasing new datasets, did you create a Datasheet for the Dataset? 
\answerNA{We are not releasing new datasets.}
\end{enumerate}

\item Additionally, if you used crowdsourcing or conducted research with human subjects, \textbf{without compromising anonymity}...
\begin{enumerate}
  \item Did you include the full text of instructions given to participants and screenshots?
    \answerYes{Yes, survey instrument is included in Appendix~\ref{app:instrument}}
  \item Did you describe any potential participant risks, with mentions of Institutional Review Board (IRB) approvals?
    \answerYes{Yes, survey instrument is included in Appendix~\ref{app:instrument}}
  \item Did you include the estimated hourly wage paid to participants and the total amount spent on participant compensation?
    \answerYes{Yes, compensation is described in \sect\ref{sec:methods_survey}.}
   \item Did you discuss how data is stored, shared, and deidentified?
   \answerYes{Yes, privacy is detailed in \sect\ref{sec:ethics}.}
\end{enumerate}

\end{enumerate}

\appendix
\onecolumn
\section{Survey Instrument}\label{app:instrument}

The following survey instrument was presented to respondents using Qualtrics.

\subsection*{Intro and Screener Questions}
The information collected in this study will be used as part of a research study conducted by  Reddit, Inc. Your survey responses will be de-identified. Any data collected will be used and shared in accordance with our Privacy Policy.

By agreeing to participate in this study, you consent to our collection and processing of your survey responses. You represent you are 18 years of age or older. You acknowledge that any information about the study, including any study details and any information provided to you through the study, are considered Reddit Confidential Information. You acknowledge Reddit may use your de-identified survey answers for marketing purposes. 

By clicking the link to the survey, you acknowledge that you have read and agree to this consent.

$\circ$ I acknowledge that I have read the consent language above and agree to participate in this survey study.

What is your age?

\begin{itemize}
\item[$\circ$] Under 15 [terminate]
\item[$\circ$] 16-17 [terminate]
\item[$\circ$] 18-24
\item[$\circ$] 25-34
\item[$\circ$] 35-44
\item[$\circ$] 45-54
\item[$\circ$] 55-64
\item[$\circ$] 65+ 
\end{itemize}

\subsection*{Reddit Usage}

How long have you been active on Reddit? [Choose the answer that best applies]

\begin{itemize}
\item[$\circ$] Less than 1 week
\item[$\circ$] 1-4 weeks
\item[$\circ$] 1-3 months
\item[$\circ$] 4-6 months
\item[$\circ$] 7 months - 1 year
\item[$\circ$] Over 1 year
\end{itemize}

How often do you typically visit Reddit? [Choose the answer that best applies]

\begin{itemize}
\item[$\circ$] Multiple times per day
\item[$\circ$] Once per day
\item[$\circ$] A few times per week
\item[$\circ$] A few times per month
\item[$\circ$] Once a month or less often
\end{itemize}

\subsection*{Subreddit-Specific Questions}

Please think about the reasons that you visit r/\underline{\hspace{2cm}} and rate your level of agreement or disagreement with each of the following statements [5-point Likert Agree/Disagree]

\begin{itemize}
\item I use this subreddit to get information
\item I use this subreddit to learn how to do things
\item I use this subreddit to provide others with information
\item I use this subreddit to contribute to the pool of information
\item I use this subreddit to meet people with my interests
\item I use this subreddit to build relationships with others
\item I use this subreddit to learn about myself
\item I use this subreddit to gain insight into myself
\item I use this subreddit to gain prestige
\item I use this subreddit to feel important
\item I use this subreddit to receive entertaining content
\item I use this subreddit to have fun
\item I use this subreddit for relaxation or stress relief
\item I use this subreddit as a way to pass time when bored
\end{itemize}

Please rate your level of agreement or disagreement with each of the statements regarding r/\underline{\hspace{2cm}}. Choose the option that best describes how you personally feel [5-point Likert Agree/Disagree]

\begin{itemize}
\item This subreddit has quality content.
\item The content in this subreddit has variety.
\item The people in this subreddit are diverse.
\item This subreddit has trustworthy people and information.
\item Members of this subreddit engage with one another.
\item Members of this subreddit are included and able to contribute.
\item This community is an appropriate size.
\item Members of this community have input into moderator decisions.
\item This community is free of offensive or harassing behavior.
\end{itemize}

Think about how important each of the following is to your experience in r/\underline{\hspace{2cm}}, and rank them in terms of importance.

\begin{itemize}
\item Quality of the content
\item Variety in/of the content
\item Diversity of the people
\item Trustworthiness of the people and information
\item Members’ engagement with one another
\item Members’ inclusion and ability to contribute
\item Size of the community
\item Community input into moderator decisions
\item Absence of offensive or harassing behavior
\item Other (please explain) \fbox{\phantom{xx years old.}}
\end{itemize}

Please rate your level of agreement or disagreement with each of the statements regarding the community within r/\underline{\hspace{2cm}}. Choose the option that best describes how you personally feel [5-point Likert Agree/Disagree]

\begin{itemize}
\item I expect to be a part of this community for a long time.
\item I think this community is a good thing for me to be a part of.
\item It is important to me to be a part of this community.
\item I feel at home in this community.
\item I recognize the screen names of most participants in this community.
\item If there is a problem in this community, members can get it solved.
\item Members of this community can be counted on to help others. 
\item I want the same things from this community as other members.
\item Members of this community share the same values. 
\item I have friends in this community that I can depend on. 
\item If I have a personal problem, I can turn to members of this community.
\item I care about what other community members think of me.
\item Most members of this community know me. 
\item I feel like I have influence over what this community is like
\end{itemize}

What other, existing subreddits would you recommend to a redditor who enjoyed and participated in r/\underline{\hspace{2cm}}, and why? (Optional)

\noindent
\fbox{\phantom{Free-Text Response, Optional}}

As it exists right now, what are a few of the best aspects of r/\underline{\hspace{2cm}}? (Optional)

\noindent
\fbox{\phantom{Free-Text Response, Optional}}

If you could change anything about r/\underline{\hspace{2cm}}, what are some aspects that could be improved upon? (Optional)

\noindent
\fbox{\phantom{Free-Text Response, Optional}}

\subsection*{Demographics and Conclusion}

How would you describe your gender identity? Please mark any answers that apply:

\begin{itemize}
\item[$\circ$] Man
\item[$\circ$] Woman
\item[$\circ$] Non-binary
\item[$\circ$] Prefer to self-describe (please specify)  \fbox{\phantom{gender response}}
\item[$\circ$] Prefer not to answer
\end{itemize}

If you’d like to receive a gratuity for completing this survey, please provide a valid email address, where you would prefer to receive your Amazon gift card. (optional)

\noindent
\fbox{\phantom{email address here}}

Is there anything else that you’d like to tell us about your experience using Reddit, in general? (optional)

\noindent
\fbox{\phantom{Free-Text Response, Optional}}

\section{Conversational Patterns}\label{app:patterns}
We computed the number of occurrences of eight unique conversational patterns in each target subreddit. While not all patterns were included in our final model (\sect\ref{sec:results_modeling}), we include them here for completeness.

\begin{table}[h]
\centering
\begin{tabular}{ll}
\textbf{Pattern Abbreviation} & \textbf{Description of Pattern}           \\ \hline
\texttt{ApBc}                 & OP Posts, Receives Reply                  \\
\texttt{ApBcAr}               & OP Posts, Receives Reply, Replies Back    \\
\texttt{ApBcAu}               & OP Posts, Receives Reply, Upvotes         \\
\texttt{ApBcAd}               & OP Posts, Receives Reply, Downvotes       \\
\texttt{AcBr}                 & OP Comments, Receives Reply               \\
\texttt{AcBrAc}               & OP Comments, Receives Reply, Replies Back \\
\texttt{AcBrAu}               & OP Comments, Receives Reply, Upvotes      \\
\texttt{AcBrAd}               & OP Comments, Receives Reply, Downvotes   
\end{tabular}
\caption{Descriptions of the eight conversational patterns that we identified occurrences of in target subreddits. Not all patterns were included in our final model.}
\label{tab:all_patterns}
\end{table}

\clearpage
\section{SOVC Scores by Topic}\label{app:topic}
\new{
To assess how SOVC varies between communities of different topics, we made use of Reddit's publicly available topic listings, which consist of 29 top level topic categories viewable at \texttt{www.reddit.com/explore}. For communities assigned multiple topics, we used only the topic assigned the highest relevance to that community. We excluded 6 topics (Nature \& Outdoors, Sciences, Adult Content, Wellness, Q\&As \& Stories, and Spooky) for which we had fewer than three participating communities assigned that topic.
}

\begin{figure}[h]
    \centering
    \includegraphics[width=\textwidth]{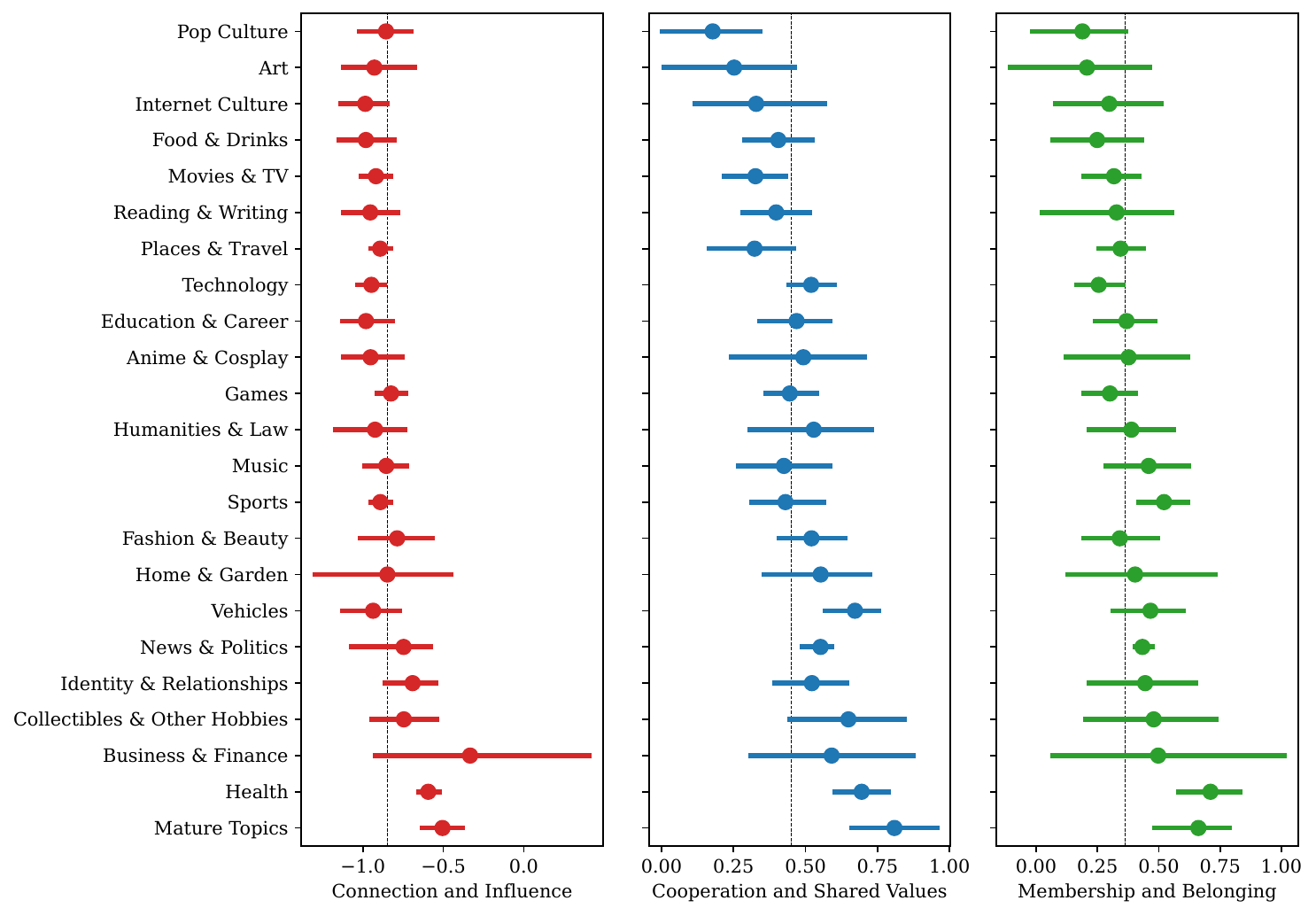}
    \caption{Communities with different topics differ somewhat in their senses of virtual community. The dotted line shows the average SOVC score across all communities included in our study, while points show the average SOVC score for communities of that type, along with bootstrapped 95\% confidence intervals. Sports communities have fairly typical scores for Connection and Influence and Cooperation and Shared Values, yet above average scores for Membership and Belonging. Health communities have above average scores for all three SOVC factors.}
    \label{fig:sovc_by_topic}
\end{figure}

\end{document}